\documentclass[useAMS,usenatbib]{mn2e}

\usepackage[]{color,graphicx}

\usepackage{amsmath}
\usepackage{subfigure}

\def\lesssim{\mathrel{\hbox{\rlap{\hbox{\lower5pt\hbox{$\sim$}}}\hbox{$<$}}}}
\def\gtrsim{\mathrel{\hbox{\rlap{\hbox{\lower5pt\hbox{$\sim$}}}\hbox{$>$}}}}

\title[Anisotropic neutrino effect on magnetar spin]
{Anisotropic neutrino effect on magnetar spin: constraint on inner toroidal field}
\author[Yudai Suwa and Teruaki Enoto]{
Yudai Suwa$^{1}$\thanks{E-mail: suwa@yukawa.kyoto-u.ac.jp}
and
Teruaki Enoto$^{2,3}$\thanks{E-mail: teru.enoto@riken.jp}
\\
$^{1}$Yukawa Institute for Theoretical Physics, Kyoto University,
Oiwake-cho, Kitashirakawa, Sakyo-ku, Kyoto, 606-8502, Japan\\
$^{2}$High Energy Astrophysics Laboratory, Institute of
  Physical and Chemical Research (RIKEN), Wako, Saitama, 351-0198,
  Japan\\
$^{3}$Goddard Space Flight Center, NASA, Greenbelt,
  Maryland, 20771, USA
}

\begin{document}

\date{Accepted. Received.}

\pagerange{\pageref{firstpage}--\pageref{lastpage}} \pubyear{2014}

\maketitle

\label{firstpage}

\begin{abstract}
The ultra-strong magnetic field of magnetars modifies the neutrino
cross section due to the parity violation of the weak interaction and
can induce asymmetric propagation of neutrinos.  Such an anisotropic
neutrino radiation transfers not only the linear momentum of a neutron
star but also the angular momentum, if a strong toroidal field is
embedded inside the stellar interior. As such, the hidden toroidal
field implied by recent observations potentially affects the
rotational spin evolution of new-born magnetars. We analytically solve
the transport equation for neutrinos and evaluate the degree of
anisotropy that causes the magnetar to spin-up or spin-down during the
early neutrino cooling phase. Supposing that after the neutrino
cooling phase the dominant process causing the magnetar spin-down is
the canonical magnetic dipole radiation, we compare the solution with
the observed present rotational periods of anomalous X-ray pulsars 1E
1841-045 and 1E 2259+586, whose poloidal (dipole) fields are $\sim
10^{15}$ G and $10^{14}$ G, respectively. Combining with the supernova
remnant age associated with these magnetars, the present evaluation
implies a rough constraint of global (average) toroidal field strength
at $B^\phi\lesssim 10^{15}$ G.
\end{abstract}


%%%%%%%%%%%%%%%%%%%%%%%%
\section{Introduction}
\label{sec:intro}
%%%%%%%%%%%%%%%%%%%%%%%%

Soft Gamma Repeaters (SGRs) and Anomalous X-ray Pulsars (AXPs) are two
examples of the astronomical objects collectively known as
magnetars. These objects emit a large amount of energy in soft gamma
rays and X-rays, and their energy source cannot be explained in terms
of the canonical rotation energy of neutron stars (NSs).  Magnetic
fields inside and outside magnetars are conjectured to be the main
source of energy, with very strong magnetic fields required to explain
their activity.\footnote{Another possible source is the accretion
  mechanism \citep[see e.g.][]{true10}, but here we concentrate on the
  strong magnetic field hypothesis in this paper. } Magnetars are
therefore a special class of NSs that have strong magnetic
fields. Based on their periods ($P$) and the time derivative of their
periods ($\dot P$), this class is thought to have magnetic fields
larger than the critical strength $B_{Q}\approx 4.4\times 10^{13}$ G,
beyond which the perturbative approach of quantum-electro dynamics
breaks down.

Recently, two magnetars with surface dipole magnetic fields smaller
than $B_{Q}$ were reported \citep{rea10,rea12}. These objects gave us
important clues as to the nature of the magnetic field inside
magnetars.  Since $P$ and $\dot P$ measurements can only provide
information on the dipole (poloidal) component of the field, there is
no constraint on the toroidal component. As such, the unknown toroidal
fields are often thought to provide the large energy required to
account for magnetar activity. The two low-magnetic field SGRs are
thought to be explained by hidden internal magnetic fields (e.g., SGR
0418+5729, \citealt{tien13}).

It is often discussed in the literature that parity violation in weak
interactions can lead to asymmetric neutrino emission in strongly
magnetized NSs.  Given that neutrinos transfer momentum, asymmetric
neutrino emission originating from poloidal fields can therefore
impart linear momentum to a NS, which is a possible cause of pulsar
kicks \citep{arra99a,ando03b,kota05,maru12}.  Furthermore, asymmetric
neutrino emission could also transfer angular momentum from new-born
NSs \citep{maru14}.

In this paper we investigate the effect of a magnetic field on the
opacity of NSs to the neutrinos that carry away the thermal energy. We
specifically focus on the toroidal component and the spin evolution of
magnetars. Section \ref{sec:scenario} opens with the basic picture of
this paper. Section \ref{sec:transfer} is devoted to the derivation of
the neutrino transfer equation and its solution. In addition, we give
simple relations between the total angular momentum of a NS and the
angular momentum emitted by neutrinos.  In Section
\ref{sec:constraint} we give the constraint on the magnetar's internal
field. We summarize our results and discuss their implications in
Section \ref{sec:summary}.

%%%%%%%%%%%%%%%%%%%%%%%%
\section{Physical Scenario}
\label{sec:scenario}
%%%%%%%%%%%%%%%%%%%%%%%%

In this section we briefly outline the basic picture studied in this
paper. As is well known, NSs are formed by the gravitational collapse
of massive stars, leading to core-collapse supernova explosions. At
first, just after their formation, NSs are hot (the temperature is
typically $O(10^{11})$ K), and in this phase they are referred to as
protoneutron stars (PNSs).  The stars then proceed to cool down due to
neutrino emission \citep[see e.g.][]{burr86,fisc10,suwa14}. The
typical timescale of the cooling, referred to as the Kelvin-Helmholtz
cooling time and denoted $\tau_\nu$ in the following, is $O(1)$
s.\footnote{This is determined by $E_\mathrm{th}/L_{\nu}$, where
  $E_\mathrm{th}$ is the thermal energy stored in the PNS and
  $L_{\nu}$ is the neutrino luminosity.}  In this paper, we are
focusing on this early PNS cooling phase. Note that this is different
from conventional NS cooling, the timescale of which is typically of
$O(10^{5})$ years.

During the PNS cooling phase, the strong magnetic field induces
anisotropic interactions between neutrinos and polarized nucleons and
electrons. These interactions lead to an anisotropic deformation of
the neutrino flux, which in turn imparts a linear momentum to the PNS
and produces a pulsar kick (Section \ref{sec:intro}).  The emitted
neutrinos may also transfer angular momentum, causing the PNS to
spin-up/down. These linear and angular momentum transfers are caused
by the strong poloidal and toroidal components of magnetic fields,
respectively. A quantitative evaluation of the angular momentum allows
us to determine the dependency of the NS spin on the toroidal field
strength.  The optical depth of neutrinos during this period is much
higher than unity, so the neutrino transfer is approximated with the
diffusion equation as derived and solved in Section
\ref{sec:transfer}. Using this solution, we give an estimate for the
angular momentum transferred as a result of the anisotropic neutrino
emission in the strong toroidal magnetic field.

\begin{figure}
\centering \includegraphics[width=0.45\textwidth]{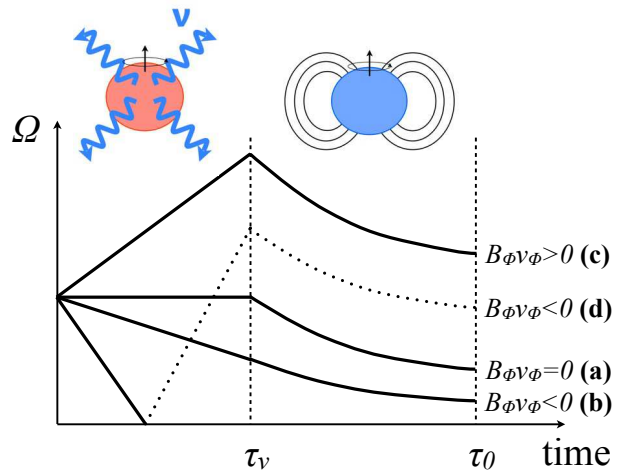} 
\caption{ Schematic view of the time evolution of angular velocity,
  $\Omega$. For $t<\tau_\nu$ the neutrino emission changes the NS spin
  and for $\tau_\nu<t<\tau_0$ the NS rotation is decelerated by the
  usual dipole radiation.  Depending on the direction of magnetic
  fields, the NS spin evolution can be classified as following.
  In the case (a), since the toroidal filed is absent, for
  $t<\tau_\nu$ the rotation velocity is not altered by neutrino
  emission;
  In the case (b), the neutrinos decelerate the NS spin;
  In the case (c), the neutrinos accelerate the NS spin;
  In the case (d), the neutrinos first decelerate the NS spin and
  eventually the NS rotation is stopped. Since the neutrinos transfer
  the angular momentum even after the NS rotation stops, then the NS
  starts counterrotating (dotted line).
  The spin deceleration by dipole radiation does not depend on the
  rotation direction, so that the spin evolution for $<\tau_\nu$ is
  similar independent on the evolution for $t<\tau_\nu$. It is clear
  that the rotation period of NSs distribute broadly if the neutrinos
  significantly affect the spin evolution. Therefore, if the neutrino
  effect dominates the spin evolution of NSs for $t<\tau_\nu$, in
  order to concentrate the current spin period of NSs in a narrow
  range, neutrino effect upon the NS spin should be small enough.  }
\label{fig:t-omega}
\end{figure}

Anisotropic neutrino emission makes the PNS slower or faster depending
on the directions of the rotation and magnetic fields during the PNS
cooling phase. After this initial phase, the magnetar spins-down due
to the canonical dipole radiation in the typical time scale of the
current pulsar age, $\tau_0$ ($\tau_{\nu} < t < \tau_0$). Let us here
consider the constraint on the toroidal magnetic field by utilizing
available present observations of magnetar spin periods.  Observed
rotational periods of magnetars are slow and localized to a narrow
range, from $\sim$2 to $\sim$11 s (see Table \ref{tab}). This means
that the total angular momentum transferred by the neutrinos in the
PNS phase is smaller than the initial NS angular momentum at that
time. If this were not the case, a fine tuning would be needed to
produce the slow spin concentration, because the direction of neutrino
angular momentum transfer does not depend on the spin direction (see
Figure \ref{fig:t-omega}). For example, if the magnitude of the
neutrino momentum transfer is larger than the initial angular
momentum, even NS spin-up is possible via momentum transfer in the
opposite direction (see case (d) in Figure \ref{fig:t-omega}).  As
such, the assumption that the transferred angular momentum is smaller
than that of the NS at $t=\tau_{\nu}$ seems reasonable.
Using the associated supernova remnant (SNR) age as the current age of
magnetars ($\tau_{0}$), we can evaluate the spin period at
$t=\tau_{\nu}$ by turning back the spin using the dipole radiation
model (see Appendix \ref{sec:appendix}). The amount of angular
momentum that can be transferred by the neutrinos can be constrained
using the angular momentum at $t=\tau_{\nu}$.  By using this
constraint we will then put an upper limit on the internal toroidal
magnetic field (see Eqs. \ref{eq:bphi} and \ref{eq:bphi2}).

%%%%%%%%%%%%%%%%%%%%%%%%
\section{Anisotropic neutrino flux and momentum transfer}
\label{sec:transfer}
%%%%%%%%%%%%%%%%%%%%%%%%

%%%
\subsection{Neutrino transfer equation}
\label{sec:equation}
%%%

Following \cite{arra99a,arra99b}, we solve the transfer equation for
neutrinos.
The Boltzmann equation for neutrinos is given by
\begin{equation}
\frac{1}{c}\frac{\partial f_\nu(\vec{p}_{\nu})}{\partial t}+\vec{\Omega}\cdot\nabla f_\nu(\vec{p}_{\nu})=S,
\label{eq:boltz}
\end{equation}
where $c$ is the speed of light, $f_{\nu}(\vec{p}_{\nu})$ is the
distribution function for neutrinos with momentum $\vec{p}_{\nu}$, $t$
is time, $\vec{\Omega}$ is the propagation direction of neutrinos, and
$S$ is the source term, in which scattering and absorption are
included.

Since we are considering the neutrino transfer inside a PNS, where the
neutrinos propagate diffusely, we employ the following diffusion
approximation for the neutrino distribution function,
\begin{equation}
f_\nu(\vec{p}_{\nu})=f_\nu^{(0)}(\epsilon_{\nu})+g(\epsilon_{\nu})+3\vec{\Omega}\cdot \vec{h}(\epsilon_{\nu}),
\end{equation}
where $f^{(0)}_{\nu}$ is the Fermi-Dirac distribution function for
neutrinos, $\epsilon_{\nu}$ is the neutrino energy, $g(\epsilon_{\nu})$
is the deviation from thermal equilibrium and
$\vec{h}(\epsilon_{\nu})$ is the dipole component that is connected to
the neutrino flux.

By averaging Eq. (\ref{eq:boltz}) over the whole solid angle and omitting the time derivative
term, we get the following moment equation for steady state
\citep{arra99a}
\begin{equation}
\nabla\left[f_\nu^{(0)}+g\right]+\epsilon_\mathrm{abs}\kappa_0^\mathrm{abs}g\hat B=-3\kappa_0^{\mathrm{tot}}\vec{h},
\label{eq:0th}
\end{equation}
where $\epsilon_\mathrm{abs}$ is a coefficient related to absorption
and originates from the existence of strong magnetic fields (if there
are no magnetic fields $\epsilon_\mathrm{abs}$ is
zero). $\kappa^\mathrm{abs}_{0}$ is the inverse of the mean free path
for neutrino emission and absorption ($p+e^{-}\rightleftharpoons
n+\nu_{e}$) and $\kappa^\mathrm{tot}_{0}$ is the inverse of the mean
free path for all interactions, including isoenergetic scattering by
nucleons without magnetic fields. Lastly, $\hat
B\equiv\vec{B}/|\vec{B}|$.

Similarly, we obtain the first order moment equation by integrating
Eq. (\ref{eq:boltz}) multiplied by $\mu=\vec{\Omega}\cdot
\vec{r}/|\vec{r}|$ as
\begin{equation}
\nabla\cdot\vec{h}=-\kappa_0^\mathrm{abs}g-\epsilon_\mathrm{abs}\kappa_0^\mathrm{abs}\vec{h}\cdot\hat B.
\label{eq:1st}
\end{equation}
Note that to obtain Eqs. (\ref{eq:0th}) and (\ref{eq:1st}) we omitted
source terms relating to the scattering originating from the existence
of magnetic fields (denoted $\epsilon_\mathrm{sc}$ in
\citealt{arra99a, arra99b}).  This is because this contribution is
much smaller that from the terms proportional to
$\epsilon_\mathrm{abs}$.\footnote{In \cite{arra99b}, they found that
  $\epsilon_\mathrm{sc}\sim
  10^{-2}\epsilon_\mathrm{abs}(e)(kT/1~\mathrm{MeV})^{-1}(\epsilon_\nu/1~\mathrm{MeV})^2$
  (see equations 7.1 and 7.2 in their paper), where
  $\epsilon_\mathrm{abs}(e)$ is the asymmetry coefficient for neutrino
  absorption by electrons, $k$ is Boltzmann's constant and $T$ is the
  matter temperature. Since we are interested in the region where
  $kT\sim\epsilon_\nu\sim O(1)$ MeV, omitting $\epsilon_\mathrm{sc}$
  is a reasonable approximation.}

Combining Eqs. (\ref{eq:0th}) and (\ref{eq:1st}), we get the following
diffusion equation
\begin{equation}
\frac{1}{3r^{2}}\frac{\partial}{\partial r}
\left[
\frac{r^{2}}{\kappa_0^\mathrm{tot}}
\frac{\partial (f_0+g)}{\partial r}
\right]
=
\kappa_0^\mathrm{abs}g.
\end{equation}
Note that we omitted the higher-order term proportional to
$\epsilon_\mathrm{abs}^{2}$.  Using the specified opacities for
$\kappa^\mathrm{abs}_{0}$ and $\kappa^\mathrm{tot}_{0}$, we can solve
this diffusion equation.

Following \citep{arra99b}, the opacities are estimated as:
\begin{align}
\kappa^\mathrm{abs}_{0}(\epsilon_{\nu})=&\frac{(G_{F}\hbar c)^{2}}{\pi}(\epsilon_{\nu}+Q)^{2} n_{n}(c_{V}^{2}+3c_{A}^{2})\left[1-f_{e}(\epsilon_{\nu}+Q)\right]\nonumber\\
=&3.66\times 10^{-9}~\mathrm{cm}^{-1}\left(\frac{\epsilon_{\nu}+Q}{2.29~\mathrm{MeV}}\right)^{2}\left(\frac{\rho}{10^{11}~\mathrm{g~cm^{-3}}}\right)\nonumber\\
&\times\left[1-f_{e}(\epsilon_{\nu}+Q)\right],\\
\kappa^\mathrm{sc}_{0}(\epsilon_{\nu})&=\frac{2}{3\pi}(G_{F}\hbar c)^{2}\epsilon_{\nu}^{2}\left(c_{V}^{2}+5c_{A}^{2}\right)n\nonumber\\
&=3.38\times 10^{-10}~\mathrm{cm}^{-1}\left(\frac{\epsilon_{\nu}}{1~\mathrm{MeV}}\right)^{2}\left(\frac{\rho}{10^{11}~\mathrm{g~cm^{-3}}}\right),\\
\kappa^\mathrm{tot}_{0}(\epsilon_{\nu})&=\kappa^\mathrm{abs}_{0}(\epsilon_{\nu})+\kappa^\mathrm{sc}_{0}(\epsilon_{\nu}).
\end{align}
Here, $G_{F}=1.166\times 10^{-5}$ GeV$^{-2}$ is Fermi's constant,
$\hbar=1.054\times10^{-27}$ cm$^{2}$ g s$^{-1}$ is the reduced Planck
constant, $Q=1.29$ MeV is the difference in mass between a neutron and
proton, $n_{n}$ is the number density of neutrons, $c_{V}$ and $c_{A}$
are weak interaction constants,\footnote{For $\nu n\to \nu n$,
  $c_{V}=-1/2$ and $c_{A}=-1.23/2$.  For $\nu p\to\nu p$,
  $c_{V}=1/2-2\sin^{2}\theta_{w}=0.035$ and $c_{A}=1.23/2$, where
  $\theta_{w}$ is the Weinberg angle.} $f_{e}$ is the distribution
function for electrons and $n$ is the number density of nucleons. For
deriving typical values we used $n_{n}=n_{p}=n/2$, where $n_{p}$ is
the number density of protons. The composition is assumed to be
completely dissociated to free protons and neutrons. We have neglected
stimulated absorption effects for simplicity.

The absorption coefficient, as given by \cite{arra99b}, is
\begin{align}
\epsilon_\mathrm{abs}
&=\frac{1}{2}\frac{(\hbar c)^{2}eB}{(\epsilon_{\nu}+Q)^{2}}\frac{c_{V}^{2}-c_{A}^{2}}{c_{V}^{2}+3c_{A}^{2}}\\
&=-0.0575
\left(\frac{B}{10^{15}~\mathrm{G}}\right)
\left(\frac{\epsilon_{\nu}+Q}{2.29~\mathrm{MeV}}\right)^{-2},
\end{align}
where $c_{V}=1$, and $c_{A}=1.26$ for absorption.

The density profile employed in this study, which mimics the structure
of the protoneutron star, is
\begin{equation}
\rho(r)=\rho_{\nu}\left(\frac{r}{R_{\nu}}\right)^{-3},
\end{equation}
where $\rho_{\nu}$ is the density of the PNS surface and $R_{\nu}$ is
the radius of the protoneutron star. Here we take $\rho_{\nu}=10^{11}$
g cm$^{-3}$ and $R_{\nu}=100$ km.\footnote{For simplicity, we neglect
  the time evolution of $R_{\nu}$, which evolves from $\sim 100$ km to
  $\sim 10$ km within the PNS cooling time.}  Although the density
diverges at the center, it does not matter in this study because
neutrinos are tightly coupled with matter and $f_{\nu}=f^{(0)}_{\nu}$
there.

By assuming that the matter temperature is constant and neutrinos are
not degenerated (i.e. taking the chemical potential of neutrinos to be
vanishing),\footnote{The temperature above the neutrinosphere, which
  we are considering in this paper, can be approximated as almost
  constant and the chemical potential of electrons is negligible
  \citep[see][]{jank01}.} we obtain the following steady state
equation for $G\equiv g/f^{(0)}_{\nu}$
\begin{equation}
G''+\frac{5}{r}G'-\alpha\left(\frac{R_{\nu}}{r}\right)^{6}G=0,
\label{eq:g}
\end{equation}
where a prime denotes the derivative with respect to $r$ and
\begin{align}
\alpha=&
4.01\times 10^{-17}\mathrm{cm}^{-2}
\left(\frac{\epsilon_{\nu}+Q}{2.29~\mathrm{MeV}}\right)^{4}
\left(1-f_{e}\right)^{2}\nonumber\\
&+
3.71\times 10^{-18}\mathrm{cm}^{-2}
\left(\frac{\epsilon_{\nu}+Q}{2.29~\mathrm{MeV}}\right)^{2}
\left(\frac{\epsilon_{\nu}}{1~\mathrm{MeV}}\right)^{2}\nonumber\\
&\times\left(1-f_{e}\right).
\end{align}
The solution to Eq. (\ref{eq:g}) is given by
\begin{equation}
G=C_{1}\frac{I_{1}(\sqrt{\alpha}R_{\nu}^{3}/2r^{2})}{r^{2}}+C_{2}\frac{K_{1}(\sqrt{\alpha}R_{\nu}^{3}/2r^{2})}{r^{2}},
\end{equation}
where $I$ and $K$ denote modified Bessel functions of the first and
second kind, respectively, and $C_{1}$ and $C_{2}$ are constants. At
the center, neutrinos are tightly coupled with matter so that
$f_{\nu}=f^{(0)}_{\nu}$ and $g=0$, meaning that $C_{1}=0$. From
Eq. (\ref{eq:0th}), the flux is given as
\begin{equation}
\vec{h}=-\frac{1}{3\kappa_{0}^\mathrm{tot}}\left(G'f_{\nu}^{(0)}\hat r+\epsilon_\mathrm{abs}\kappa_0^\mathrm{abs}Gf_{\nu}^{(0)}\hat B\right),
\end{equation}
where $\hat r$ denotes the unit vector in the radial direction. Since
the specific neutrino flux is given by
$\vec{F}_{\nu}=(\epsilon_{\nu}/2\pi\hbar c)^{3}c\vec{h}$, $r-$ and
$\phi-$components are given as
\begin{align}
F^{r}_{\nu}&=-\frac{c}{3\kappa_{0}^\mathrm{tot}}\left(\frac{\epsilon_{\nu}}{2\pi\hbar c}\right)^{3}\left(G'+\epsilon_\mathrm{abs}\kappa_0^\mathrm{abs}G\frac{B^{r}}{B}\right)f_{\nu}^{(0)},
\label{eq:fr}\\
F^{\phi}_{\nu}&=-\frac{c}{3\kappa_{0}^\mathrm{tot}}\left(\frac{\epsilon_{\nu}}{2\pi\hbar c}\right)^{3}\epsilon_\mathrm{abs}\kappa_{0}^\mathrm{abs}G\frac{B^{\phi}}{B}f^{(0)}_{\nu}.
\end{align}
Here, $B^{r}$ and $B^{\phi}$ correspond to the $r-$ and
$\phi-$components of the magnetic field, respectively. $F^{r}_{\nu}$
should be positive at $R_{\nu}$ so that $C_{2}<0$.

By integrating over energy, using the matter temperature $kT=4$ MeV
and vanishing chemical potentials for $f^{(0)}_{\nu}$ and $f_{e}$, the
ratio between fluxes in the radial and orthogonal directions at the
neutrinosphere surface is given by
\begin{equation}
\left.\frac{\int d\epsilon_{\nu} F^{\phi}_{\nu}}{\int d\epsilon_{\nu} F^{r}_{\nu}}\right|_{r=R_{\nu}}
\approx -0.013
\left(\frac{B^{\phi}}{10^{15}~\mathrm{G}}\right)
\left(\frac{R_\nu}{100~\mathrm{km}}\right)^{1/2}.
\label{eq:asym}
\end{equation}
The second term in Eq. (\ref{eq:fr}) is neglected in this estimation.

The total neutrino luminosity is given by
\begin{equation}
L_{\nu}=\int d\epsilon_{\nu}d\Omega F^{r}_{\nu} R_{\nu}^{2},
\label{eq:lnu}
\end{equation}
and the rate of angular momentum transfer by neutrinos is given by
\begin{equation}
J_{\nu}=\frac{1}{c}\int d\epsilon_{\nu}d\Omega F^{\phi}_{\nu}R_{\nu}^{3}\sin\theta.
\label{eq:jnu}
\end{equation}
The factor $R_{\nu}\sin\theta$ comes from the distance from the
symmetry axis.
By combining Eqs. (\ref{eq:asym}), (\ref{eq:lnu}) and (\ref{eq:jnu}),
and assuming that $F^{r}_{\nu}$ is independent of the angle, we obtain
\begin{align}
J_{\nu}
=&
-0.013
\left(\frac{\left<B^{\phi}\right>}{10^{15}~\mathrm{G}}\right)
\left(\frac{R_\nu}{100~\mathrm{km}}\right)^{1/2}
\frac{R_{\nu}L_{\nu}}{c}\\
=&
-4.3\times 10^{47}~\mathrm{g~cm^2~s^{-2}}\nonumber\\
&\times
\left(\frac{\left<B^{\phi}\right>}{10^{15}~\mathrm{G}}\right)
\left(\frac{R_\nu}{100~\mathrm{km}}\right)^{3/2}
\left(\frac{L_\nu}{10^{53}~\mathrm{erg~s^{-1}}}\right),
\label{eq:jnu2}
\end{align}
where $\left<B^{\phi}\right>\equiv \int d\Omega
B^{\phi}\sin\theta/4\pi$, which is the angle-averaged strength.

%%%
\subsection{Angular momentum transfer by neutrinos}
\label{sec:momentum}
%%%

In this subsection we evaluate the angular momentum transferred by the
anisotropic neutrino radiation that interacts with the toroidal
magnetic field. This process occurs during the PNS cooling phase when
the neutrino diffusion approximation is valid in the stellar interior
(Section \ref{sec:equation}).  By comparing it with the total angular
momentum of a rotating NS, we are able to determine an expression for
the critical magnetic field strength at which the NS rotation period
is drastically affected by the anisotropic neutrino radiation. In
order to compare with present observations, here we employ the NS
angular momentum at a stellar radius of 10 km after the PNS cooling
phase.  This assumption is valid if the angular momentum is conserved
when the PNS (i.e. hot NS) contracts to a cold NS, where the radius
shrinks from $\sim$100 km to $\sim$10 km.

The angular momentum of a NS is written as
\begin{align}
\mathcal{M}_\mathrm{NS}^{\phi}
&= I\Omega\nonumber\\
&= 7.0\times 10^{45}~\mathrm{g~cm^{2}~s^{-1}}
\left(\frac{P}{1~\mathrm s}\right)^{-1}
\left(\frac{M}{1.4M_{\odot}}\right)
\left(\frac{R_\mathrm{NS}}{10~\mathrm{km}}\right)^{2},
\label{eq:m_pns}
\end{align}
where $I=\frac{2}{5}MR^{2}_\mathrm{NS}$ is the moment of inertia,
$\Omega$ is the angular velocity, $P$ is the rotation period
($P=2\pi/\Omega$), $M$ is the NS mass and $R_\mathrm{NS}$ is the NS
radius.

The angular momentum transferred by neutrino radiation is given by
\begin{align}
\mathcal{M}_{\nu}^{\phi}&= \beta \frac{R_{\nu}E_{\nu}}{c}\nonumber\\
&= 6.7\times10^{48}~\mathrm{g~cm^{2}~s^{-1}}
\beta
\left(\frac{R_{\nu}}{100~\mathrm{km}}\right)
\left(\frac{E_{\nu}}{2\times 10^{52}~\mathrm{erg}}\right),
\label{eq:m_nu}
\end{align}
where $\beta$ is the asymmetry parameter for neutrino emission and
$E_{\nu}$ is the total energy emitted by the neutrinos responsible for
the change in spin, which is related to the luminosity as
$E_{\nu}=\int dt L_{\nu}$.  Note that a PNS has larger radius than an
ordinary NS due to the existence of thermal pressure \citep[see
  e.g.][]{jank12,suwa13b}. Although the total amount of energy that
can be released by the neutrinos is $\sim 3\times 10^{53}$ erg, the
contributions from $\nu_{\mu}$ ($\nu_{\tau}$) and $\bar\nu_{\mu}$
($\bar\nu_{\tau}$) to the change in spin cancel each other
\citep{arra99b}. As such, we only consider the energy released due to
the $\nu_{e}$ emitted in electron capture ($p+e\to n+\nu_{e}$) just
after the core bounce of supernova shock, which is $\sim O(10^{52})$
erg. The total number of $\nu_{e}$ emitted due to electron capture is
estimated as
\begin{equation}
N_{\nu_{e}}=N_{p}=\frac{MY_{p}}{m_{p}}=
8.3\times 10^{56}
\left(\frac{M}{1.4M_{\odot}}\right)
\left(\frac{Y_{p}}{0.5}\right),
\end{equation}
where $N_{p}$ is the total number of protons in the neutron star,
$m_{p}$ is the proton mass and $Y_{p}$ is the proton fraction. By
taking the average energy of emitted $\nu_{e}$ to be 3.15$kT=12.6$ MeV
($kT$/4 MeV), the total energy released due to $\nu_{e}$ emission in
the neutralization process is given as $E_{\nu_{e}}=1.7\times
10^{52}~\mathrm{erg}(M/1.4M_{\odot})(Y_{p}/0.5)(kT/4~\mathrm{MeV})$.\footnote{Note
  that, due to the difference in number density of neutrons and
  protons, the distribution functions of $\nu_{e}$ and $\bar\nu_{e}$
  may be different, meaning that the contributions from these species
  to the change in spin may not exactly cancel. In this case,
  $E_{\nu}$ could be $\sim 10^{53}$ erg, which should be checked using
  a more sophisticated neutrino transfer calculation.}

Comparing Eqs. (\ref{eq:m_pns}) and (\ref{eq:m_nu}), one recognizes
that the slowly rotating ($P\sim 1$ s) PNS's rotation can be
significantly affected if $\beta\sim 10^{-3}$. This condition can be
used to put a constraint on the strength of internal toroidal magnetic
fields. From Eqs. (\ref{eq:jnu2}) and (\ref{eq:m_nu}), $\beta$ is
given as
\begin{equation}
\beta\approx-0.013
\left(\frac{\left<B^{\phi}\right>}{10^{15}~\mathrm{G}}\right)
\left(\frac{R_\nu}{100~\mathrm{km}}\right)^{1/2},
\label{eq:beta}
\end{equation}
where we have used $\int dt L_\nu=E_\nu$. Using these relations, in
the next section we will constrain the internal toroidal field.

%%%%%%%%%%%%%%%%%%%%%%%%
\section{Constraint on Internal Toroidal Fields}
\label{sec:constraint}
%%%%%%%%%%%%%%%%%%%%%%%%

It is natural to expect that the angular momentum transferred by
neutrinos should be smaller than the total angular momentum of the PNS
at $t=\tau_\nu$ (see Section \ref{sec:scenario}).  As such, using
Eqs. (\ref{eq:m_pns}) and (\ref{eq:m_nu}) we get the following
constraint:
\begin{align}
|\beta|
\lesssim
&1.0\times 10^{-3}
\left(\frac{P}{1~\mathrm s}\right)^{-1}
\left(\frac{M}{1.4M_{\odot}}\right)
\left(\frac{R_\mathrm{NS}}{10~\mathrm{km}}\right)^{2}\nonumber\\
&\times\left(\frac{R_{\nu}}{100~\mathrm{km}}\right)^{-1}
\left(\frac{E_{\nu}}{2\times 10^{52}~\mathrm{erg}}\right)^{-1},
\end{align}
which can be rewritten as a constraint on the magnetic fields
using Eq. (\ref{eq:beta}) as
\begin{align}
\left|\left<B^{\phi}\right>\right|
\lesssim
&8.1\times 10^{13}~\mathrm{G}
\left(\frac{P}{1~\mathrm s}\right)^{-1}
\left(\frac{M}{1.4M_{\odot}}\right)
\left(\frac{R_\mathrm{NS}}{10~\mathrm{km}}\right)^{2}\nonumber\\
&\times\left(\frac{R_{\nu}}{100~\mathrm{km}}\right)^{-3/2}
\left(\frac{E_{\nu}}{2\times 10^{52}~\mathrm{erg}}\right)^{-1}.
\label{eq:bphi}
\end{align}
By exploiting the fact that the magnetic flux is conserved during the
PNS cooling phase, i.e.
$\left<B_\mathrm{NS}^{\phi}\right>R_\mathrm{NS}^2=\left<B^{\phi}\right>R_\nu^2$,
we can evaluate the field strength inside a {\it cold} NS whose radius
is $R_\mathrm{NS}$ as
\begin{align}
\left|\left<B_\mathrm{NS}^{\phi}\right>\right|
\lesssim
&8.1\times 10^{15}~\mathrm{G}
\left(\frac{P}{1~\mathrm s}\right)^{-1}
\left(\frac{M}{1.4M_{\odot}}\right)\nonumber\\
&\times\left(\frac{R_{\nu}}{100~\mathrm{km}}\right)^{1/2}
\left(\frac{E_{\nu}}{2\times 10^{52}~\mathrm{erg}}\right)^{-1}.
\label{eq:bphi}
\end{align}
We therefore see that the constraint on the magnetic field strength
depends on the rotation period $P$ at $t=\tau_\nu$.
The typical spin period of magnetars at $t=\tau_\nu$ is unclear due to
the lack of knowledge on magnetar formation.  However, if we take
$P=10$ ms at $t=\tau_\nu$, we obtain
$\left|\left<B_\mathrm{NS}^{\phi}\right>\right|\lesssim 10^{18}$ G.

If we assume that magnetic dipole radiation is the dominant process
affecting magnetar spin evolution for $t>\tau_\nu$,\footnote{Here we
  assume that the spin evolution induced by anisotropic neutrino
  radiation ceases at $t=\tau_{\nu}$ ($\sim O(1)$ s). After that only
  the long-term ($\sim 1$ kyr) spin evolution due to dipole radiation
  is considered. This is because at $t=\tau_{\nu}$ the average energy
  of neutrinos decreases and the NS becomes transparent to them, so
  that the mechanism investigated in this study is no longer active.}
the spin period of 1E~1841-045 at $t=\tau_{\nu}$ can be estimated as
$\approx$ 8-11 s (see Appendix \ref{sec:appendix}).  Therefore, using
Eq. (\ref{eq:bphi}), we can obtain the following constraint on the
field strength:
\begin{equation}
\left|\left<B_\mathrm{NS}^{\phi}\right>\right|
\lesssim 10^{15}~\mathrm{G}
\left(\frac{R_{\nu}}{100~\mathrm{km}}\right)^{1/2},
\label{eq:bphi2}
\end{equation}
where we have employed canonical values for $M$ and $E_{\nu}$. A
similar value is obtained for the case of 1E
2259+586.\footnote{Interestingly, this value is similar to the recent
  observational suggestion by \citet{maki14}, which is based on the
  pulse modulation analysis implying the precession. Note that their
  employed magnetar is different one from ours so that this
  coincidence might be just a product of chance.}  Thus, the toroidal
magnetic fields of these magnetars can be comparable to the dipole
component at least at the moment of birth.  Note that this constraint
only applies to the global toroidal field, i.e. the angle averaged
value near the NS surface, since the angular momenta transferred by
turbulent components on small scales cancel each other out.

%%%%%%%%%%%%%%%%%%%%%%%%
\section{Summary and Discussion}
\label{sec:summary}
%%%%%%%%%%%%%%%%%%%%%%%%

In this paper we studied the spin evolution of magnetars resulting
from the anisotropic neutrino emission induced by strong magnetic
fields.  We solved the diffusion equation for neutrinos and estimated
the degree of anisotropy. By considering the toroidal component of the
magnetic fields we were able to constrain the unseen internal fields
using the current rotation period of magnetars. Supposing that the
associated SNR age is the real magnetar age, we found the constraint
$\left|\left<B^{\phi}_\mathrm{NS}\right>\right| \lesssim 10^{15}$ G
for 1E1841-045 and 1E 2259+586, whose dipole fields are thought to be
$\sim 10^{15}$ G and $10^{14}$ G, respectively.

In addition to the spin evolution, we can also estimate the pulsar
kick velocity of magnetars using Eq. (\ref{eq:asym}). When we consider
the split monopole poloidal field at the PNS surface, the degree of
asymmetry $\gamma$ is $O(10^{-2})(B_{p}/10^{15}~\mathrm{G})$. The kick
velocity can thus be estimated as
\begin{align}
v_\mathrm{kick}
&=\gamma\frac{E_{\nu} }{Mc}\\
&\approx
24.0~\mathrm{km~s^{-1}}
\left(\frac{\gamma}{10^{-2}}\right)
\left(\frac{E_{\nu}}{2\times 10^{52}~\mathrm{erg}}\right)
\left(\frac{M}{1.4 M_{\odot}}\right)^{-1}.
\end{align}
We therefore see that the magnetar kick resulting from this mechanism
is expected to be very small.

In this paper we focused on magnetars (SGRs and AXPs). However, there
are other classes of stars that also have strong dipole fields
\cite[see][for a list]{dall12}. These objects exhibit a similar spin
period to magnetars (3 s $\lesssim P\lesssim $ 11 s), but their
magnetic fields are typically weaker. Even though they do not have
associated SNR, we can apply the same analysis as discussed in this
paper taking $P_i\sim O(1)$ s.  Thus, the constraint obtained in this
study is applicable for these objects as well as magnetars.

To finish we comment on the assumptions made in this study. 
First, we employed the diffusion approximation for the neutrino
radiative transfer equation. This assumption is essentially valid for
the region of the magnetar considered in this work, but near the
surface, where the mean free path of neutrinos is comparable to the
scale size, this approximation starts to break down. However, since we
are considering the region inside the PNS, the effect of the
break-down of this assumption is not significant.
Secondly, for simplicity we have assumed that the PNS radius is
constant during the cooling phase. However, this assumption does not
change our discussion drastically because the constraints on the
internal toroidal magnetic field given by Eqs. (\ref{eq:bphi}) and
(\ref{eq:bphi2}) imply very weak dependence on the PNS radius. In
addition, since a smaller PNS radius gives a tighter upper limit for
the toroidal field, our assumption of constant radius will tend to
give more conservative upper limits.
Thirdly, since the real age of a magnetar is unknown, we assumed it to
be the same as that of the SNR. Because the SNR age contains systemic
errors, this approximation might affect the derived constraint.
However, we expect that the corrections to the age do not change it by
orders of magnitude, meaning that our discussion in the previous
section should not change very much even if we include this systematic
error.
Finally, we have assumed that after neutrino emission the sole
mechanism behind the magnetar spin-down is dipole radiation. There are
several other mechanisms that can decelerate a NS's spin \citep[see
  e.g.][]{thom04}, which will tend to lead to looser constraints on
the internal fields.  This is because these mechanisms usually act
later than the neutrinos so that a smaller $P_i$ is possible. More
detailed studies that include the effects of other deceleration
mechanisms are necessary. A fundamental limit can be obtained using
the fastest rotation of a NS (i.e. the rotational breakup speed),
which gives $\left<B^\phi_\mathrm{NS}\right>\lesssim10^{19}$ G.

\section*{Acknowledgements}

We thank the referee, U. Geppert, for providing constructive comments
and help in improving the contents of this paper. YS would like to
thank P. Cerda-Duran and N. Yasutake for informative discussions,
K. Hotokezaka, T. Muranushi, and M. Suwa for comments, and J. White
for proofreading.  We also thank the Yukawa Institute for Theoretical
Physics at Kyoto University, where part of this work was done during
the workshop YITP-T-13-04 entitled ``Long-term Workshop on Supernovae
and Gamma-Ray Bursts 2013''. YS is supported in part by Grant-in-Aid
for Scientific Research on Innovative Areas (No. 25103511) and by HPCI
Strategic Program of Japanese MEXT. TE is supported by JSPS KAKENHI,
Grant-in-Aid for JSPS Fellows, 24-3320.

%%%%%%%%%%%%%%%%%%%%%%%%%%%%%%%%%%%%%%%%%%%%
\appendix
%%%%%%%%%%%%%%%%%%%%%%%%
\section{Spin evolution of magnetars}
\label{sec:appendix}
%%%%%%%%%%%%%%%%%%%%%%%%

%%%
\subsection{Case without magnetic field decay}

Since the real age of a magnetar, $\tau_{0}$, is unknown, the
characteristic spin-down time, $\tau_{c}\equiv P/2\dot P$, is
conventionally used as an approximation.  We also know that some
magnetars can be associated with SNRs, for which alternative, better
age estimations are possible via X-ray plasma diagnostics.  Here we
assume that the SNR age is a better estimator of $\tau_0$, and
extrapolate the current rotation period to the initial period at
$\tau_{\nu}$ using the dipole radiation model.
In the following discussion we give expressions for the initial
rotation period $P_i$ at $\tau_{\nu}$ and its evolution.  In this
subsection we neglect the magnetic field decay, which will be
discussed in the next subsection.

When dipole radiation is the leading cause of spin-down, the rotation
period as a function of time, $t$, can be written as \citep{shap83}
\begin{equation}
P=P_i\left(1+\frac{2P^2}{P_i^2}\frac{t}{T}\right)^{1/2},
\label{eq:P}
\end{equation}
where the initial period, $P_i$, at $t=\tau_{\nu}$ is given at the
time when dipole radiation becomes the dominant process for spin down
and
\begin{align}
T=&\frac{P}{\dot P}=\frac{3Ic^{3}P^{2}}{2\pi^{2}B_{p}^{2}R^{6}\sin^{2}\alpha} \label{eq:T}\\
=&145~\mathrm{years}
\left(\frac{B_p}{10^{15}~\mathrm{G}}\right)^{-2}
\left(\frac{R}{10~\mathrm{km}}\right)^{-4}
\left(\frac{M}{1.4M_\odot}\right)
\left(\frac{P}{1~\mathrm{s}}\right)^2,
\end{align}
where $B_p$ is the surface dipole field at the pole. Here we employ
$\sin^{2}\alpha=1$ for simplicity. Using this relation we find
\begin{align}
B_{p}&=
\left(\frac{3Ic^{3}}{2\pi^{2} R^{6}}P\dot P\right)^{1/2}\nonumber\\
&=6.75\times 10^{19}~\mathrm{G}
\left(\frac{M}{1.4M_\odot}\right)
\left(\frac{R}{10~\mathrm{km}}\right)^{-4}
\left(\frac{P}{1~\mathrm{s}}\right)^{1/2}
\left(\frac{\dot P}{1~\mathrm{s/s}}\right)^{1/2}.
\label{eq:bp}
\end{align}
Although this result looks different by a factor of two to the
frequently used $B=3.2\times 10^{19}~\mathrm{G}\sqrt{P\dot P}$, this
difference just comes from a difference in notation.\footnote{In this
  paper we use the value of the magnetic field at the pole as opposed
  to the value in the equatorial plane that is often used.} By
substituting Eq. (\ref{eq:T}) into ({\ref{eq:P}}), we get the
following simple form as
\begin{equation}
P^2=P_i^2+\frac{4\pi^{2}B_{p}^{2}R^{6}\sin^{2}\alpha}{3Ic^{3}}t.
\end{equation}

\begin{figure}
\centering \includegraphics[width=0.45\textwidth]{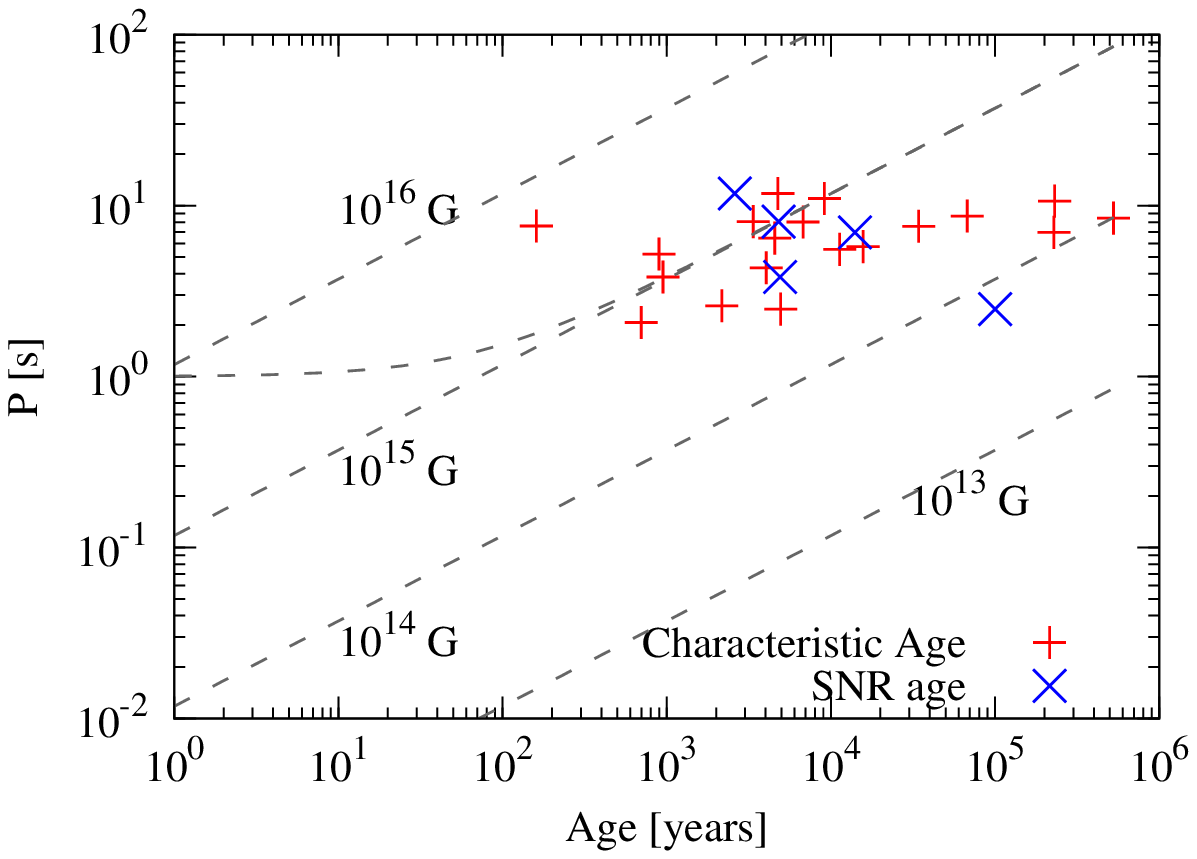} 
\caption{The time evolution of rotation period for NSs with different
  dipole magnetic fields (grey dashed lines). The imposed magnetic
  field strengths are shown near the corresponding lines. Red and blue
  points indicate the observational data for which characteristic ages
  ($\tau_{c}=P/2\dot P$) and SNR ages are used, respectively. For
  $B_{p}=10^{15}$ G we plot two lines with different initial periods.
  The top and bottom lines correspond to initial periods of 1 s and 1
  ms, respectively.}
\label{fig:t-p}
\end{figure}

In Figure \ref{fig:t-p} we show the evolution of the spin period of
neutron stars with various strengths of the constant dipole field.
The red crosses correspond to observed magnetars for which the
characteristic age is used ($\tau_{c}\equiv P/2\dot P$), whilst the
blue points correspond to magnetars that can be associated with SNRs,
so that the SNR age is used.  For $B_{p}=10^{15}$ G we plot the
evolution for two different initial periods ($P_{i}$=1 s for the top
line and 1 ms for the bottom line).  One finds that the evolutions
coincide after $\gtrsim$ 1000 years, from which we conclude that
$P_{i}$ does not affect the late time evolution.

\begin{figure}
\centering \includegraphics[width=0.45\textwidth]{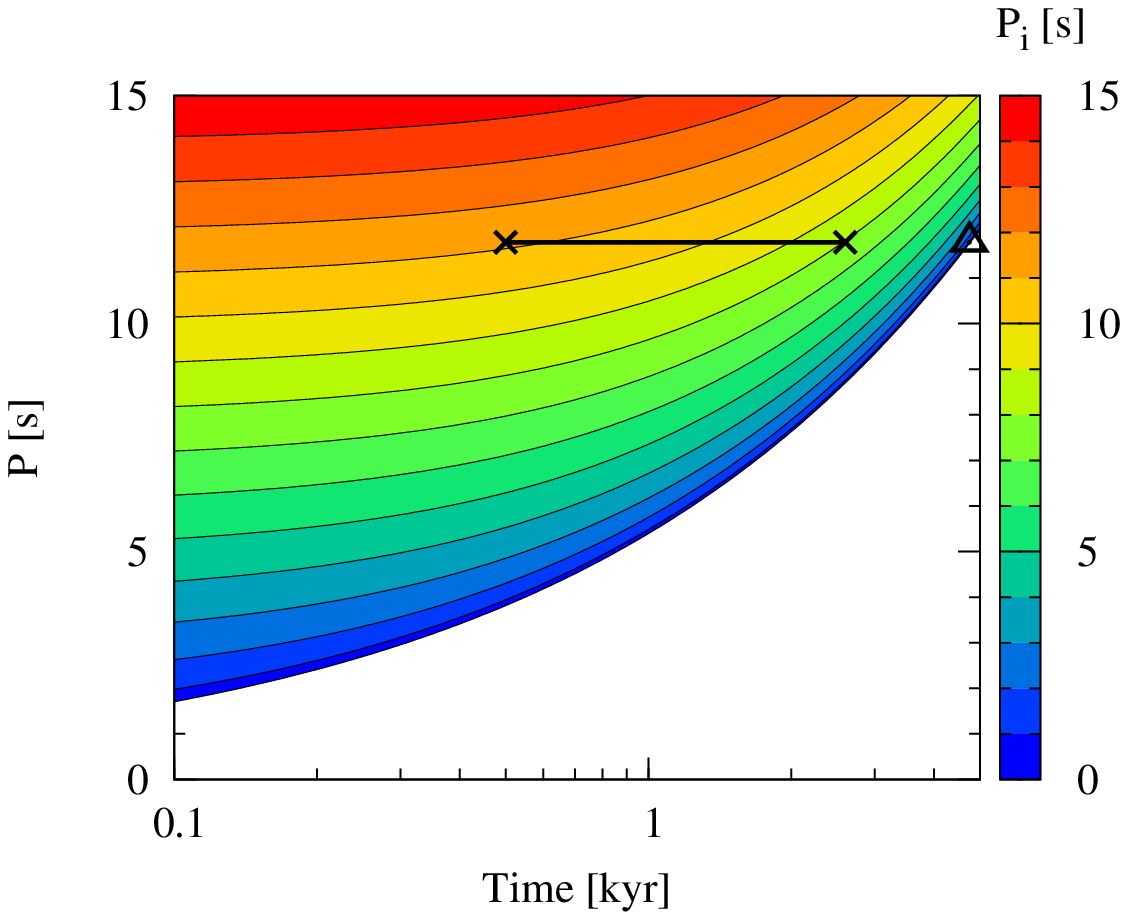}
\caption{Period evolution with time for NSs with values of $P$ and
  $\dot P$ equal to those of 1E 1841-045. The black contour lines
  correspond to trajectories with different initial spin periods,
  $P_i$.  The value of $P_i$ can be read off from the color map. The
  thick horizontal black line represents the SNR age including
  systematic errors as given in \citet{tian08}, with the two crosses
  marking the lower and upper limits of 0.5 kyr and 2.6 kyr,
  respectively.  In order to explain observational data, $P_{i}
  \approx$ 8--11 s is necessary. The triangle corresponds to the
  characteristic age ($\approx$ 4.8 kyr), and lies on a trajectory
  with infinitely small $P_{i}$.}  \label{fig:1E1841}
\end{figure}

As can be seen in Table \ref{tab}, there are two magnetars for which
the SNR age is younger than the characteristic age.  For example, 1E
2259+586 and associated SNR CTB 109 exhibit a large discrepancy
between the two ages.\footnote{In \citet{naka12} an attempt has been
  made to reconcile this discrepancy by including magnetic field
  decay.  Also note that, despite the discrepancy, it has been
  suggested that in the context of broad-band spectroscopy the
  characteristic age may be a suitable parameter to label Magnetar
  classes \citep{enot10}.} Here we treat the SNR age as the true age
and use this to estimate the spin periods of the magnetars at
birth. In Figure \ref{fig:1E1841} we show the time evolution of the
spin period for values of $P$ and $\dot P$ equal to those of 1E
1841-045. We find that $P_{i}$ should be $\approx$ 8--11 s in order to
explain the current observation with the age of $\sim$1 kyr. The same
analysis also gives the initial period of 1E 2259+586 as $P_{i}\approx
7$ s, which is almost the same as the current period.  Note that these
values would be smaller if decay of the poloidal magnetic field were
included, which will be discussed in the next subsection.

%%%
\subsection{Case with magnetic field decay}

In this subsection we study spin evolution including
phenomenologically the effect of magnetic field decay.  It is
important to consider the effect of the decaying magnetic field
because there is no isolated NS with $P\gtrsim 12$ s, meaning that the
dipole radiation can be assumed to become small enough so as to not
affect the spin period for slowly rotating NSs. There are several
studies that investigate the long-term evolution of magnetic fields
including their decay \citep[e.g.,][]{colp00,dall12,naka12,pons13}.

Using the model of \cite{colp00} and \cite{dall12}, after several
algebraic steps we get the following expressions for the time
evolution of the spin period and the dipole magnetic field strength:
\begin{align}
P^2(t)&=P_\infty^2-(P_\infty^2-P_i^2)
\left(1+\frac{t}{\tau_d}\right)^{(\alpha_B-2)/\alpha_B},\label{eq:p2}\\
B_p(t)&=\frac{B_i}{(1+t/\tau_d)^{1/\alpha_B}}\label{eq:bt},
\end{align}
where $P_\infty$ is the final spin period, $\tau_d$ is the decay
timescale of the magnetic fields, $\alpha_B$ is a parameter describing
the magnetic field decay and $B_i$ is the initial magnetic field
strength. In \cite{dall12} it was found that models with
$1.5\lesssim\alpha_B\lesssim1.8$ can explain most of the observational
evidence for isolated neutron stars with strong magnetic fields (not
only magnetars but also X-ray dim isolated NSs). Although $P_\infty$
is unknown, \cite{dall12} and \cite{pons13} suggested that
$P_\infty\approx 12$ s, because there is no observed NS with $P\gtrsim
12$ s. Thus, we employ $P_\infty=12$ s as a fiducial value here. In
addition, \cite{dall12} showed that taking $10^{15}~\mathrm{G}\lesssim
B_i\lesssim 10^{16}~\mathrm{G}$ gives good agreement with the
distribution of observed NSs with strong magnetic fields in the
$\tau_c$-$B_p$ plane.  We thus use $B_i=10^{16}$ G in the
following. In order to explain observed features, \cite{dall12}
suggested that $\tau_d=$1 kyr/$(B_i/10^{15}~\mathrm{G})^{\alpha_B}$.

\begin{figure*}
\centering \subfigure[$P_{\infty}=12$ s]{\includegraphics[width=0.45\textwidth]{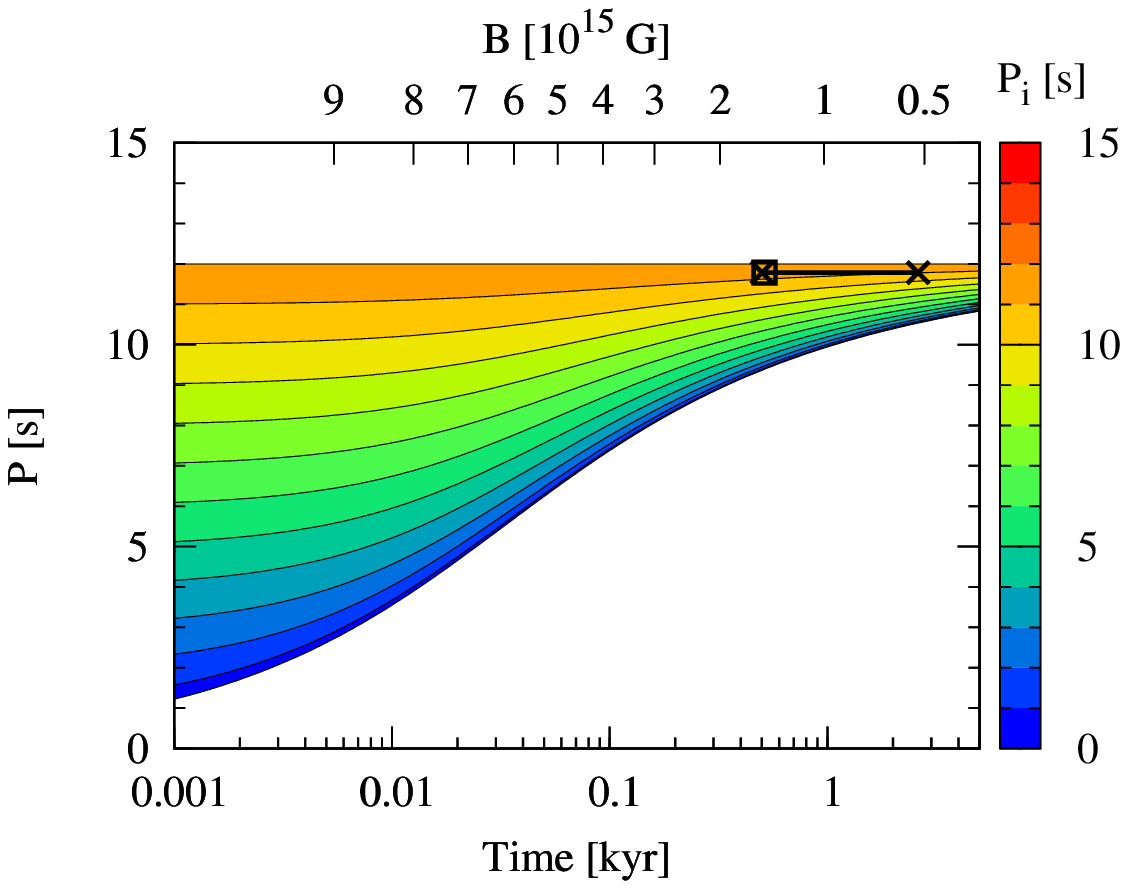}}
\subfigure[$P_{\infty}=15$ s]{\includegraphics[width=0.45\textwidth]{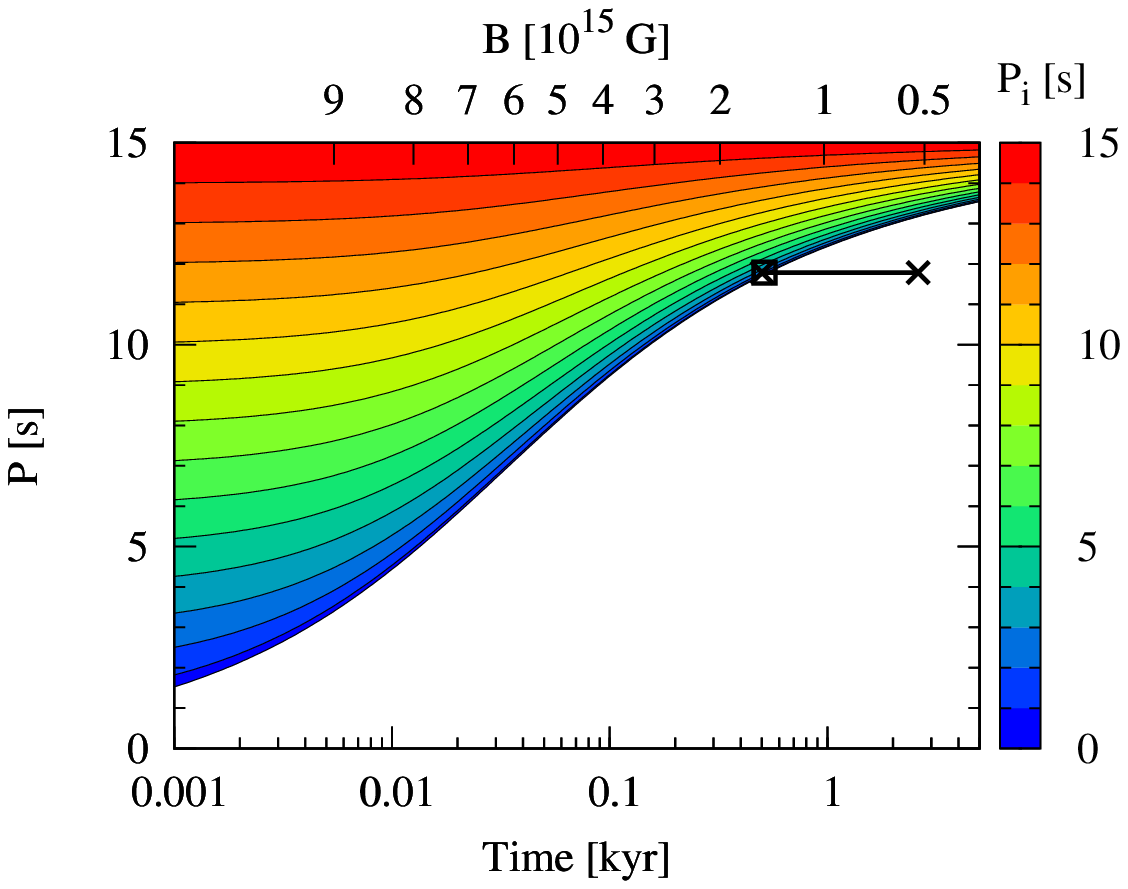}} 
\caption{ The same as Fig. \ref{fig:1E1841} but for the decaying
  magnetic field model (see Eqs. \ref{eq:p2} and \ref{eq:bt}) with the
  initial magnetic field $B_i=10^{16}$ G. The top axis corresponds to
  the strength of poloidal dipole magnetic field (see
  Eq. \ref{eq:bt}). The left panel is for $P_\infty=12$ s and the
  right panel for $P_\infty=15$ s.  The blank square marks the current
  observed $B_p$ and $P$, and is almost coincident with the left-hand
  cross that marks the lower limit on the SNR age.}
\label{fig:1E1841_p-b}
\end{figure*}

In Figure \ref{fig:1E1841_p-b} we show the period evolution of
magnetars as determined using the decaying magnetic field model. In
this figure the top axis gives the strength of poloidal field
(decreasing from the initial value of $10^{16}$ G). The blank square
shows the current position of 1E1841-045 in the $P$-$B_p$ plane, as
estimated from $P$ and $\dot P$.  We see that the square overlaps with
the left-hand cross, which corresponds to the lower limit on the SNR
age. As such, this model can be used to consistently explain all three
observed quantities $P$, $B_p$ and the SNR age.
One can see that $P_i\gtrsim 11$ s is still required in order to
explain observations using the decaying magnetic field model with
fiducial model parameters (case (a)). As such, the discussion in the
previous subsection is still valid in this case.  We do note, however,
that with a fine tuning of the parameters it is possible to explain
observational data with $P_\infty>12$ s and $P_i\ll 1$ s (see case
(b)).
On the other hand, 1E 2259+586 has $P=6.9789484460$ s. We find that
$P_i\sim 5$ s by the same discussion with fiducial parameters, which
is similar value as 1E 1841-045.  Therefore, even with the decaying
magnetic field model, we find that $P_i$ should be $O(1)$ s.

\begin{table*}
\centering
\caption{Observational properties of SGRs and AXPs}
\begin{tabular}{lccccc}
\hline
SGR/AXP name$^\dag$ & 
$P$ [s] & 
$\dot P$ [$10^{-11}$ s/s] & 
$B_p$ [$10^{14}$ G]$^\ddag$& 
$\tau_{c}$ [kyr]$^\S$
 & SNR age [kyr]\\
\hline
SGR 0418+5729          & 9.07838827(4)    & $<$0.0006   & $<$0.16 & $2.4\times10^4<$ & --- \\
SGR 0501+4516          & 5.76209653(3)    & 0.582(3)    & 3.9     & 16               & --- \\ 
SGR 0526-66            & 8.0544(2)        & 3.8(1)      & 12      & 3.4              & 4.8$^\P$ \\
SGR 1627-41            & 2.594578(6)      & 1.9(4)      & 4.7     & 2.2              & --- \\
SGR 1806-20            & 7.6022(7)        & 75(4)       & 51      & 0.16             & --- \\
Swift J1822.3-1606     & 8.43771977(4)    & 0.0254(22)  & 0.99    & 530              & --- \\
SGR 1833-0832          & 7.5654084(4)     & 0.35(3)     & 3.5     & 34               & --- \\
Swift J1834.9-0846     & 2.4823018(1)     & 0.796(12)   & 3.0     & 4.9              & 60--200$^\#$ \\ 
SGR 1900+14            & 5.19987(7)       & 9.2(4)      & 15      & 0.90             & --- \\
\hline
CXOU J010043.1-721134  & 8.020392(9)      & 1.88(8)     & 8.3     & 6.8              & ---\\ 
4U 0142+61             & 8.68832877(2)    & 0.20332(7)  & 2.8     & 68               & ---\\
1E 1048.1-5937         & 6.457875(3)      & $\sim$2.25  & 8.1     & 4.5              & ---\\
1E 1547.0-5408         & 2.0721255(1)     & $\sim$4.7   & 6.7     & 0.70             & N/A\\
PSR J1622-4950         & 4.3261(1)        & 1.7(1)      & 5.8     & 4.0              & ---\\
CXO J164710.2-455216   & 10.6106563(1)    & $\sim$0.073 & 1.9     & 230              & ---\\
1RXS J170849.0-400910  & 11.003027(1)     & 1.91(4)     & 9.8     & 9.1              & ---\\
CXOU J171405.7-381031  & 3.82535(5)       & 6.40(14)    & 11      & 0.95             & 4.9$^\%$\\
XTE J1810-197          & 5.5403537(2)     & 0.777(3)    & 4.4     & 11               & ---\\
1E 1841-045            & 11.7828977(10)   & 3.93(1)     & 15      & 4.8              & 0.5--2.6$^\&$\\
1E 2259+586            & 6.9789484460(39) & 0.048430(8) & 1.2     & 230              & 14$^\$$ \\ 
\hline
\end{tabular}
\begin{flushleft}
$^\dag$Data taken from McGill SGR/AXP Online Catalog \citep{olau14} (see also \citealt{viga13}).\\
$^\ddag$The estimation is based on Eq. (\ref{eq:bp}).\\
$^\S$Characteristic ages estimated as $P/2\dot P$.\\
$^\P$\cite{park12}.\\
$^\#$\cite{tian07}.\\
$^\%$\cite{ahar08}.\\
$^\&$\cite{tian08}.\\
$^\$$\cite{sasa13}.
\end{flushleft}
\label{tab}
\end{table*}

% for references
\newcommand\aap{{A\&A}}%                                                
          % Astronomy and Astrophysics  
\newcommand\apjl{{ApJ}}%                                                
          % Astrophysical Journal
\newcommand\apj{{ApJ}}%                                                
          % Astrophysical Journal, Letters   
\newcommand\apjs{{ApJS}}%                                                
          % Astrophysical Journal, Sapplement
\newcommand\mnras{{MNRAS}}%                                             
          % Monthly Notices of the RAS
\newcommand\prc{{Phys. Rev. C}}%                                             
          % Physical Review C
\newcommand\prd{{Phys. Rev. D}}%                                             
          % Physical Review D
\newcommand\pasj{{PASJ}}%
          % Publications of the ASJ
\newcommand\nat{{Nature}}%
          % Nature

\end{document}